\documentclass[prb,superscriptaddress,twocolumn,showpacs,floatfix,amsfonts]{revtex4}
\usepackage{graphicx,graphics,color,epsfig,exscale}% Include figure files
\usepackage{bm}
\usepackage{amsmath}
\usepackage{amssymb}

\newcommand{\crit}{\mathrm{crit}}
\newcommand{\half}{\text{$\textstyle\frac{1}{2}$}}
\renewcommand{\Im}{\text{Im}} 
\newcommand{\pdag}{\phantom{\dag}}
\newcommand{\sgn}{\text{sgn}}
\renewcommand{\Re}{\text{Re}}

\begin{document}
\title{Quantum criticality of the sub-ohmic spin-boson model}

\author{Stefan Kirchner}
\affiliation{Max Planck Institute for the Physics of Complex Systems,
N{\"o}thnitzer Str.\ 38, 01187 Dresden, Germany}
\affiliation{Max Planck Institute for Chemical Physics of Solids,
01187 Dresden, Germany}
\author{Kevin Ingersent}
\affiliation{Department of Physics, University of Florida,
P.O.\ Box 118440, Gainesville, Florida 32611--8440, USA}
\author{Qimiao Si}
\affiliation{Department of Physics \& Astronomy, Rice University, Houston,
Texas 77005, USA}

\begin{abstract}
We revisit the critical behavior of the sub-ohmic spin-boson model.
Analysis of both the leading and subleading terms in the temperature
dependence of the inverse static local spin susceptibility at the quantum
critical point, calculated using a numerical renormalization-group method,
provides evidence that the quantum critical point is interacting in cases
where the quantum-to-classical mapping would predict mean-field behavior.
The subleading term is shown to be consistent with an $\omega/T$ scaling of
the local dynamical susceptibility, as is the leading term.
The frequency and temperature dependences of the local spin susceptibility in
the strong-coupling (delocalized) regime are also presented.
We attribute the violation of the quantum-to-classical mapping to a
Berry-phase term in a continuum path-integral representation of the model.
This effect connects the behavior discussed here with its counterparts in
models with continuous spin symmetry. 
\end{abstract}


\maketitle

\section{Introduction}

The standard Kondo model, describing exchange scattering between a local moment
and a fermionic band,\cite{Hewson} can be generalized to various Bose-Fermi
Kondo models that include coupling of the impurity to one or more bosonic baths.
It is conventional to consider bosonic baths described by a power-law density
of states
\begin{equation}
\label{EQ:sub-Ohmic}
\sum_p \delta(\omega-w_p) \propto
   \omega^{1-\epsilon} ~\Theta(\omega_c-\omega) ,
\end{equation}
where $w_p$ is the dispersion of the bosonic bath (see below), $\Theta$ is the
Heaviside function, and $\omega_c$ is a high-energy cut-off. For a subset of
sub-ohmic (positive-$\epsilon$) baths corresponding to $0<\epsilon<1$, the
Bose-Fermi Kondo problem has a quantum (temperature $T=0$) phase
transition\cite{Zhu.02,Zarand.02,Glossop.07,Kirchner.09} between a Kondo phase
in which the ground state is a Kondo singlet formed between the local moment
and conduction-electron spins, and a local-moment phase in which the coupling
to the bosonic bath inhibits the Kondo effect. In all the Bose-Fermi Kondo
models that have been studied---corresponding to Ising, $XY$, and SU(2)
symmetry of the bosonic couplings---this transition is continuous. 

The Landau theory for continuous phase transitions invokes the
fluctuations of an order parameter. For the Bose-Fermi Kondo problem, a
suitable order parameter is the local magnetization $M=\langle S^z\rangle$,
which vanishes in the Kondo phase but is nonzero throughout the local-moment
phase. Within the Landau approach, the quantum criticality is described by a
local $\phi^4$ theory with a dissipative quadratic term corresponding to a
long-ranged interaction in imaginary time:
$\frac{1}{2} S^z(\tau)\chi_0^{-1}(\tau-\tau')S^z(\tau')$ with
\begin{equation}
\chi_0^{-1} (\tau-\tau') \sim \frac{1}{|\tau - \tau'|^{2 - \epsilon}} .
\label{chi-0}
\end{equation}
In keeping with the standard theory of quantum criticality, this description
is called a quantum-to-classical mapping.

In a study of the Bose-Fermi Kondo model with SU($N$) spin symmetry and SU($M$)
channel symmetry, it was shown in the limit of large $N$ and $M$ with $M/N$
fixed that the quantum critical point (QCP) breaks the quantum-to-classical
mapping.\cite{Zhu.04}
The violation is especially clear for $\half<\epsilon<1$. In this range, the
local-$\phi^4$ description gives rise to a Gaussian fixed point,\cite{Fisher.72}
and a dangerously irrelevant $\phi^4$ coupling leads to a critical static spin
susceptibility
\begin{equation}
\chi_{\crit}^{\mathrm{cl}} (T,\omega=0) \sim \frac{1}{T^x}
  \quad\text{with~~$x=\half$~~for $\half<\epsilon<1$}.
\label{chi-loc-cl}
\end{equation}
However, the large-$N$ results for the local spin susceptibility and Green's
function in the quantum critical regime of the
$\mathrm{SU}(N)\times\mathrm{SU}(M)$ Bose-Fermi Kondo model satisfy $\omega/T$
scaling, which implies that the QCP is interacting. In particular, 
\begin{equation}
\chi_{\crit}^{\mathrm{qu}} (T,\omega=0) \sim \frac{1}{T^x}
  \quad\text{with~~$x=1-\epsilon$}
\label{chi-loc-qu}
\end{equation}
and
\begin{equation}
\chi_{\crit}^{\mathrm{qu}} (T=0,\omega) \sim \frac{1}{(-i\omega)^y}
  \quad\text{with ~~$y=x$}
\label{chi-loc-qu-dyn}
\end{equation}
hold over the entire range $0<\epsilon<1$.
This violation of the quantum-to-classical mapping survives the inclusion of
$1/N$ corrections, and has been attributed to the effect of the Berry phase in
the spin path-integral representation of the local moment.\cite{Kirchner.08e}

Subsequently, a violation of quantum to classical mapping was also discussed
for the sub-ohmic spin-boson model.\cite{Vojta.05} This model (fully specified
in Section II) couples the component $S^z$ of an SU(2) local spin to the
displacement of a sub-ohmic bosonic bath with coupling constant $g$ and the
component $S^x$ to a transverse magnetic field $\Gamma$. The model has a QCP
between a delocalized phase in which the impurity degree of freedom is quenched
by the field and a boson-dominated localized phase that retains a two-fold
local-moment degree of freedom. These phases are analogous to the Kondo and
local-moment phases, respectively, of the Bose-Fermi Kondo model. In the
special case where the bosons form an ohmic bath, the spin-boson model and the
fermionic Kondo model can be transformed into one another via
bosonization\cite{Guinea.85}; however, this transformation breaks down for
$\epsilon > 0$. In Ref.\ \onlinecite{Vojta.05}, it was shown using a bosonic
extension\cite{Bulla.03,Bulla.05} of the numerical renormalization-group (NRG)
method that the critical exponents for $\half<\epsilon<1$ obey hyperscaling
(as they do for $0<\epsilon<\half$). In particular, just as for the Bose-Fermi
Kondo model,\cite{Glossop.07,Kirchner.09} the critical local susceptibility
satisfies Eqs.~\eqref{chi-loc-qu} and \eqref{chi-loc-qu-dyn}, suggestive of
an interacting fixed point with $\omega/T$ scaling---a conclusion at odds
with the quantum-to-classical mapping.

Recently, two sets of Monte-Carlo calculations\cite{Kirchner.09,Winter.09}
were carried out for a one-dimensional classical spin chain with long-range
Ising interactions $\sum_{i,j} J_{ij} S_i^z S_j^z$, where
$J_{ij} \sim 1/|r_i - r_j|^{2-\epsilon}$ for $|r_i - r_j|\gg\tau_0$.
Reference~\onlinecite{Kirchner.09} applied a cluster Monte-Carlo method at
various nonzero values of the short-range cutoff $\tau_0$, while
Ref.~\onlinecite{Winter.09} used a similar method after taking the limit
$\tau_0 \rightarrow 0$.
These works both confirmed an earlier conclusion\cite{Luijten.97}
that the critical points of such long-ranged classical spin chains are
interacting for $0<\epsilon<\half$ and Gaussian for $\half<\epsilon<1$,
consistent with the predictions of the local $\phi^4$
theory.\cite{Fisher.72,Brezin.82}

The results of the classical Monte-Carlo calculations have been interpreted in
two very different ways. In Ref.~\onlinecite{Kirchner.09}, we suggested that
these results differ from NRG calculations for the Bose-Fermi Kondo model
because a classical spin chain of length $L$ sites does not faithfully
represent the quantum-mechanical model at temperature $T=1/(L\tau_0)$.
More specifically, the limit $\tau_0 \rightarrow 0$ of the classical spin
chain does not reproduce the path integral of the quantum problem because
this limit smears the topological effect of the Kondo spin flips.
Along this line, it has recently been shown \cite{Kirchner.10} that a proper
path-integral for the sub-ohmic spin-boson model involves a Berry-phase term.
Numerically, Ref.~\onlinecite{Kirchner.09}, generalizing the NRG result of
Ref.~\onlinecite{Glossop.07}, showed that the temperature dependence of the
local spin susceptibility of the Bose-Fermi Kondo model obeys
Eq.\ \eqref{chi-loc-qu} over about 20 decades in temperature.

By contrast, Ref.~\onlinecite{Winter.09} assumed the the classical spin chain
faithfully represents the spin-boson model. Correspondingly, it interpreted the
Monte-Carlo result as demonstrating a fundamental error in the NRG
results\cite{Bulla.03,Bulla.05,Vojta.05} for the sub-ohmic spin-boson model.
It was further suggested in Refs.\ \onlinecite{Vojta.09} and
\onlinecite{Vojta.10} that over the range
$\half<\epsilon<1$, the NRG result for the temperature dependence of the local
spin susceptibility obeys Eq.\ \eqref{chi-loc-qu} due to an artifact of the
method,
reminiscent of the effect of a temperature-dependent mass term \cite{note_mass_winding}
discussed in
Ref.~\onlinecite{Kirchner.09},
and that removal of a spurious $T^{1-\epsilon}$
term from $\chi_{\crit}^{-1} (T)$ exposes an underlying $T^{1/2}$ term
representing the true critical behavior.

In this paper, we investigate more thoroughly the temperature and frequency
dependences of the local spin susceptibility of the spin-boson model, as
calculated using the bosonic NRG approach.~\cite{Bulla.03,Bulla.05}
The subleading term in the temperature dependence at $\omega=0$ is shown to be
described by an exponent $x_2$ that depends on $\epsilon$ and exceeds $\half$,
contradicting a central assumption of Ref.\ \onlinecite{Vojta.10}.
The exponent $y_2$ of the subleading term in the frequency dependence at $T=0$
satisfies $y_2=x_2$, paralleling the equality $y=x$ of the leading exponents. 
This provides evidence that both the leading and subleading terms in the local
susceptibility satisfy $\omega/T$ scaling. Finally, we show that NRG gives 
consistent frequency and temperature dependences of the local spin susceptibility
in the strong-coupling (delocalized) phase. All these features indicate that
the temperature and frequency dependences of the critical local spin
susceptibility are consistent with an interacting fixed point for
$\half<\epsilon<1$, thereby augmenting previous evidence for the violation
of quantum-to-classical mapping in the sub-ohmic spin-boson and Bose-Fermi
Kondo models.

The remainder of the paper is organized as follows. The model and the bosonic
NRG method are described in Sec.\ \ref{model_method}. Section \ref{chi_qcp} is
devoted to the analysis of the leading and subleading terms in the temperature
and frequency dependences of the inverse local spin susceptibility at the QCP.
In Sec.~\ref{chi_deloc}, we discuss the local susceptibility in the
strong-coupling (delocalized) phase.  Section \ref{discussion} addresses the
implications of our results, the form of the proper path integral for the
spin-boson model, and the role of the Berry phase.
The paper concludes with a brief summary in Sec.\ \ref{summary}.

\section{Model and Solution Method}
\label{model_method}

The spin-boson model is described by the Hamiltonian
\begin{equation}
\label{EQ:H-sb}
\mathcal{H}_{\text{SB}} = - \Gamma S^x
+ g \, S^z \sum_p \Bigl( \phi_{p}^{\pdag} + \phi_{-p}^{\;\dag} \Bigr)
  + \sum_p w_p \: \phi_p^{\dag} \, \phi_p^{\pdag} \, ,
\end{equation}
where $S^{\alpha} = \sigma^{\alpha}/2$ with $\sigma^{\alpha}$ being a Pauli
matrix, and $\phi_p$ annihilates a boson in a bath whose density of states is
specified by Eq.\ \eqref{EQ:sub-Ohmic}. The cutoff frequency $\omega_c$ entering
Eq.\ \eqref{EQ:sub-Ohmic} will be set to unity and will henceforth serve as
the unit of energy.

We have studied the model using the NRG method described in
Refs.\ \onlinecite{Bulla.03} and \onlinecite{Bulla.05}.
The bosonic bath is divided into a set of bins spanning oscillator frequencies
$\Lambda^{-(k+1)}<\omega<\Lambda^{-k}$ for $\Lambda>1$ and
$k = 1$, 2, 3, $\ldots$. Within each bin, the continuum of bath states is
replaced by a single state, namely, the linear combination of states that
couples to the impurity spin. Then the Lanczos method is used to map the
spin-boson Hamiltonian to
$H_{\text{SB}}^{\text{NRG}} = \lim_{n\to\infty} H_n$, where
\begin{align}
\label{EQ:H-sb-NRG}
\mathcal{H}_n
&= - \Gamma S^x + g\,S^z \sum_p \Bigl( b_0^{\pdag} + b_0^{\dag} \Bigr) \notag\\
&\quad  + \sum_{m=0}^n \Bigl[ e_m b_m^{\dag} b_m^{\pdag} +
  t_m \Bigl(b_m^{\dag} b_{m-1}^{\pdag} + \text{H.c.} \Bigr) \Bigr] .
\end{align}
As a result of the logarithmic binning, the tight-binding coefficients $e_m$
and $t_m$ that encode the bath density of states decay as $\Lambda^{-m}$.
This decay allows the Hamiltonian $H_{\text{SB}}^{\text{NRG}}$ to be solved
iteratively starting with $H_0$ and using the eigensolution of $H_n$ to
construct the basis of $H_{n+1}$. 
The basis of each site of the bosonic chain
must be truncated at $b_m^{\dag} b_m^{\dag} < N_b$. Even with restriction, the
dimension of the Fock space grows exponentially with iteration number $n$.
After a few iterations, it is possible to retain only the $N_s$ many-body
eigenstates of lowest energy after iteration $n$. All the results reported
below were obtained using $N_b = 24$ and $N_s = 300$.

Our focus in this paper is on the temperature and frequency dependences 
of the local spin susceptibility.
Within the NRG approach, the local static susceptibility is calculated as
\begin{equation}
\chi(T,\omega=0) = \lim_{h\to 0} -\frac{\langle S^z\rangle}{h}\, ,
\end{equation}
where $h$ is a local magnetic field coupling to the $S^z$ component of
the localized spin through an additional Hamiltonian term
$\Delta H = h S^z$.
The imaginary part of the local dynamical susceptibility can be computed as
\begin{align}
\chi''(T,\omega)
&= \frac{\pi}{Z_n} \sum_{j,k}|{}_n\langle k| S^z|j\rangle_n|^2
   \left(e^{-E_{n,k}/T}-e^{-E_{n,j}/T}\right) \notag \\
&\qquad \times\delta(\omega-E_{n,k}+E_{n,j}),
\end{align}
where $|j\rangle_n$ is a many-body eigenstate of iteration $n$ with energy
$E_{n,j}$, and $Z_n=\sum_j e^{-E_{n,j}/T}$.
In order to minimize known artifacts of the NRG discretization, we calculate 
the local susceptibility only at iterations of the same parity (either $n$ even
or $n$ odd) and at only one temperature or frequency per iteration:
$T_n, \omega_n \propto \Lambda^{-n}$.

\section{Local correlation functions at the quantum critical point}
\label{chi_qcp}

\begin{figure}
\centering
\includegraphics[width=0.7\linewidth]{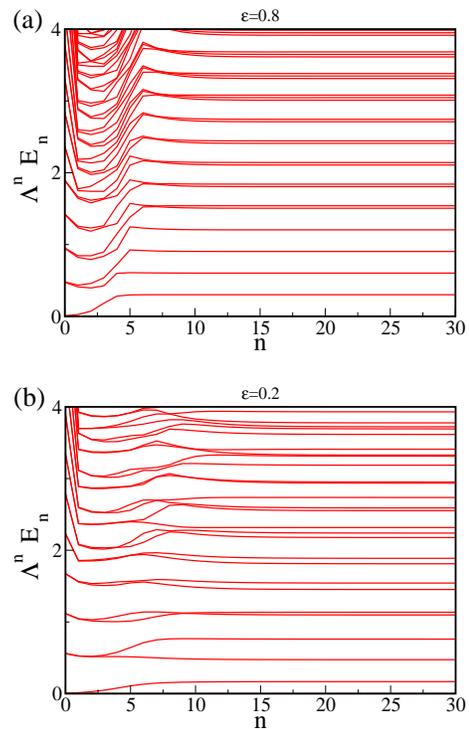}
\caption{(Color online) NRG spectra of the spin-boson model at its QCP:
Scaled NRG energy $\Lambda^n E_n$ vs iteration number $n$ for a transverse
field $\Gamma=0.01$, NRG discretization $\Lambda=3$, and bath exponent
(a) $\epsilon=0.8$, (b) $\epsilon=0.2$.
The flatness in the $n$-dependence of the scaled energy shows that the system
is at a renormalization-group fixed point; here, it corresponds to the QCP at
$g=g_c$.}
\label{nrg-spectrum}
\end{figure}

The QCP is located in our NRG calculations by fixing the transverse field
$\Gamma$ and tuning $g$. The critical strength $g_c$ is identified as the value
of $g$ at which the scaled NRG many-body energies $\Lambda^n E_n$ are
independent of the iteration number $n$, signifying that the system is at a
scale-invariant fixed point. This invariance is illustrated in
Fig.~\ref{nrg-spectrum} for two values of $\epsilon$. The
critical many-body spectrum is qualitatively similar for 
all baths corresponding to $0<\epsilon < 1$.

\subsection{Temperature dependence of the critical local susceptibility}
\label{chiT_qcp}

\begin{figure}
\centering
\includegraphics[width=0.75\linewidth]{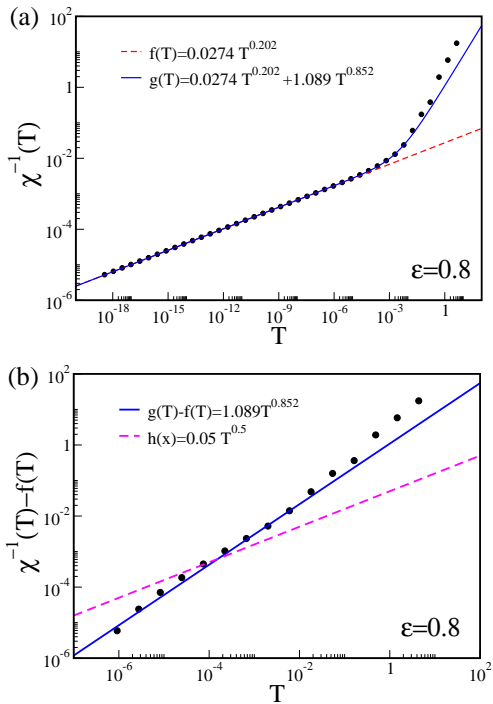}
\caption{(Color online) Temperature dependence of the local static
susceptibility at the QCP ($g=g_c$). (a) $\chi^{-1}(T)$ for
$\epsilon=0.8$, $\Gamma=0.01$, and $\Lambda=3$. The dashed line represents
a fit of $\chi^{-1}(T)$ to the leading term $f(T)$ specified in
Eq.\ \eqref{chi-fit-leading}, which yields an exponent $x=0.202$. The solid
line is a fit to Eq.\ \eqref{chi-fit-leading-and-subleading} in terms of both
leading and subleading terms.
(b) Residual $\chi^{-1}(T) - f(T)$ after subtraction of the fitted leading term
from (a). The fitted solid line gives a subleading exponent $x_2=0.852$. The
residual clearly does not have a $T^{1/2}$ dependence (dashed line).}
\label{chi-T-eps-8}
\end{figure}

\begin{figure}
\centering
\includegraphics[width=0.75\linewidth]{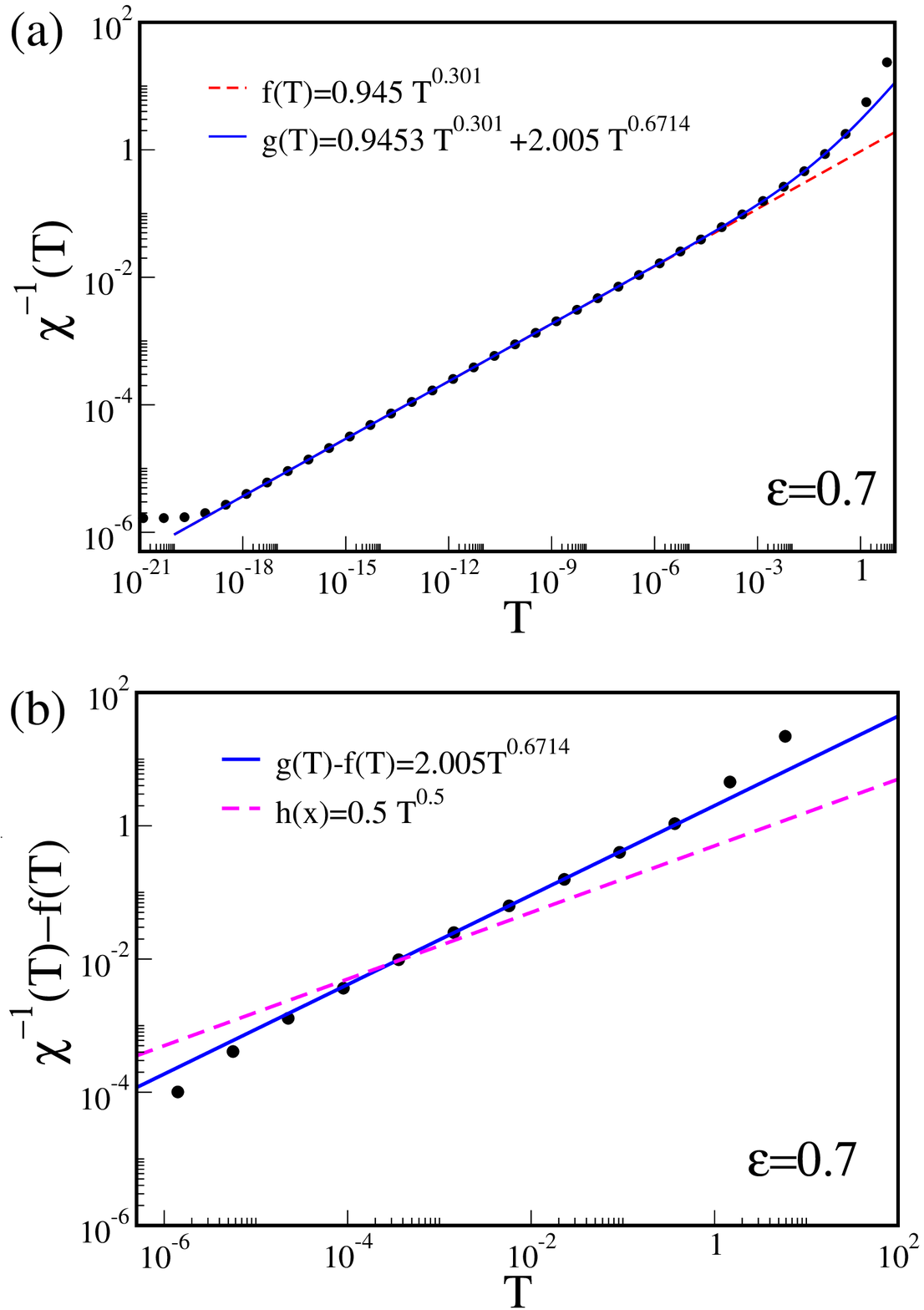}
\caption{(Color online) Like Fig.\ \ref{chi-T-eps-8} but for
$\epsilon=0.7$, $\Gamma=1$, and $\Lambda=4$.}
\label{chi-T-eps-7}
\end{figure}

\begin{figure}
\centering
\includegraphics[width=0.75\linewidth]{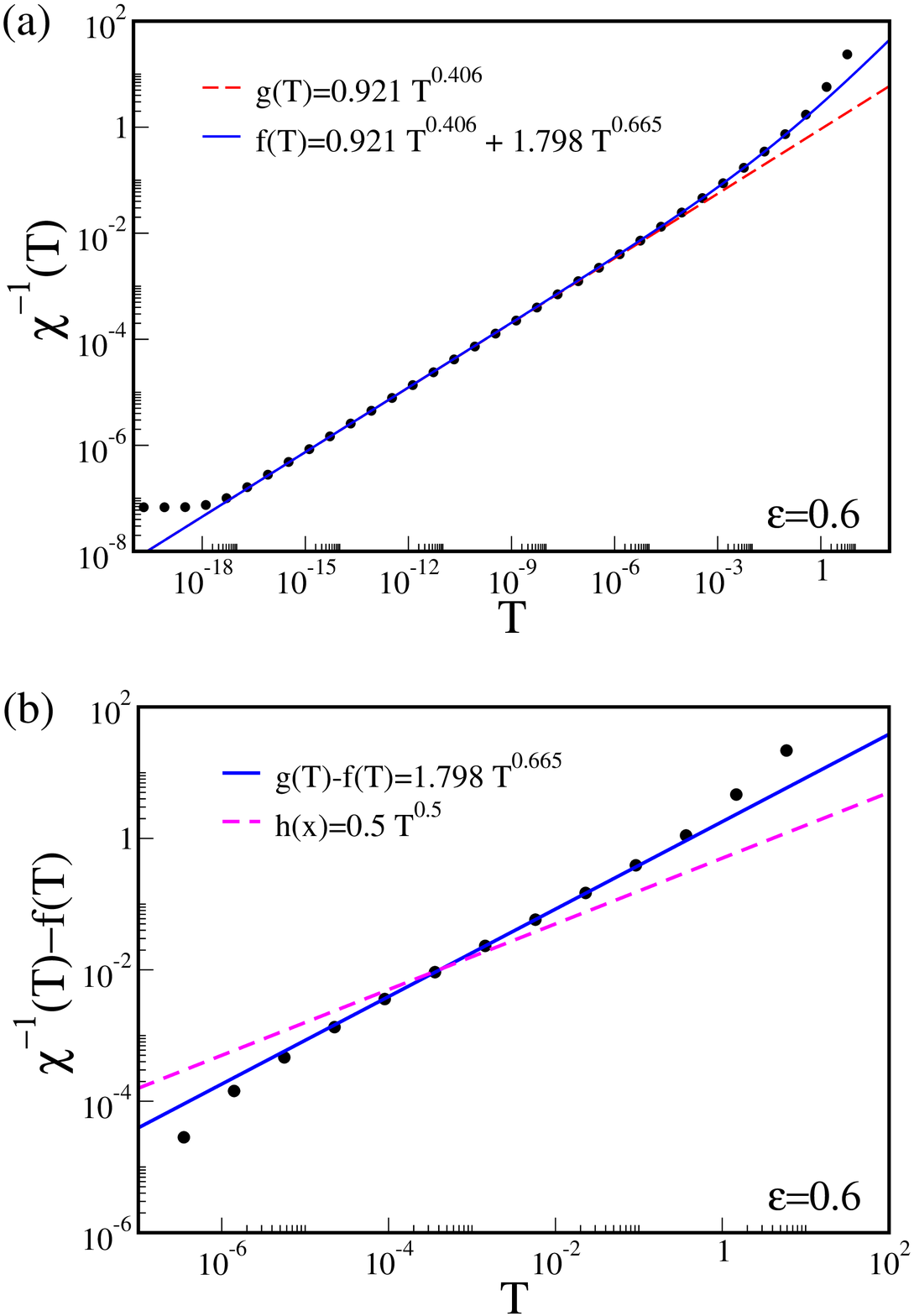}
\caption{(Color online) Like Fig.\ \ref{chi-T-eps-7}, but for
$\epsilon=0.6$.}
\label{chi-T-eps-6}
\end{figure}

\begin{figure}
\centering
\includegraphics[width=0.75\linewidth]{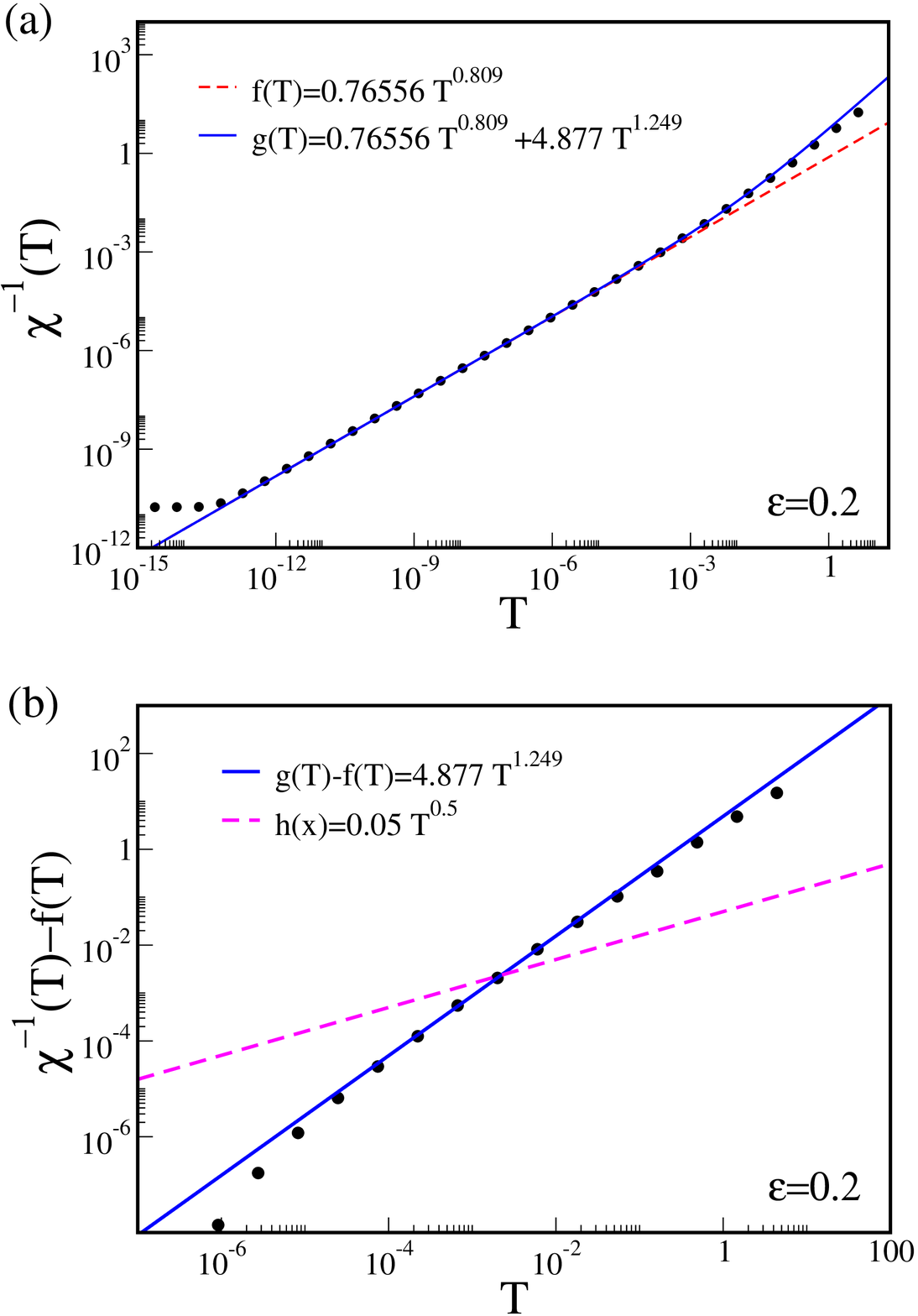}
\caption{(Color online) Like Fig.\ \ref{chi-T-eps-8}, but for
$\epsilon=0.2$.}
\label{chi-T-eps-2}
\end{figure}

At the QCP, $g=g_c$, the local static spin susceptibility has a singular
temperature dependence. Figure \ref{chi-T-eps-8}(a) is a log-log plot of
$\chi^{-1}$ vs $T$ for $\epsilon=0.8$.
A least-squares fit (dashed line) of the logarithm of $\chi^{-1}$ to the
logarithm of a simple power law
\begin{equation}
f(T)= a T^x
\label{chi-fit-leading}
\end{equation}
over $10^{-18}<T<10^{-4}$ yields a critical exponent of $x=0.202$. 
Similar behavior is observed in the local static susceptibility
for $\epsilon=0.7$, fitted over $10^{-19}<T<10^{-4}$ by $x=0.301$
[Fig.\ \ref{chi-T-eps-7}(a)];
for $\epsilon=0.6$, fitted over $10^{-17}<T<10^{-4}$ by $x=0.406$
[Fig.\ \ref{chi-T-eps-6}(a)];
and for $\epsilon=0.2$, fitted over $10^{-13}<T<10^{-4}$ by $x=0.809$
[Fig.\ \ref{chi-T-eps-2}(a)].
These temperature exponents are consistent with $x=1-\epsilon$
to within about $1.5\%$.

The modest discrepancies between the exponents $x$ extracted from the
numerical data and their interacting values $1-\epsilon$ can be attributed
to a combination of fitting uncertainties and errors associated with
the NRG method. Any NRG calculation contains discretization errors
introduced by working with $\Lambda > 1$, as well as truncation errors
arising from retaining only $N_s$ many-body states after each iteration.
For the bosonic bath treated here, there is also the need to truncate
the dimension of the Fock space of each bosonic orbital on the NRG chain
to a finite value $N_b$. One should always bear in mind such sources of
systematic error, as one would for any numerical method. Still, several
features of the leading-order results serve as nontrivial checks;
in particular, the NRG value of the frequency exponent $y$ of the critical
local susceptibility (see Sec.\ \ref{chiw_qcp}) agrees with that of an
$\epsilon$ expansion\cite{Zhu.02}, and (b) the temperature exponent $x$
and several other critical exponents satisfy hyperscaling relations (as
discussed in Sec.\ \ref{discussion}).

Motivated by considerations laid out in the introduction, we further explore
the critical behavior by examining the subleading temperature
dependence defined through the fitting function
\begin{equation}
\chi^{-1}(T) = a T^x + a_2 T^{x_2} .
\label{chi-fit-leading-and-subleading}
\end{equation}
For the case $\epsilon=0.8$ shown in Fig.\ \ref{chi-T-eps-8}(a), fitting to
Eq.\ \eqref{chi-fit-leading-and-subleading} yields a subleading exponent
$x_2=0.852$. The quality of the fit of the subleading term can be seen more
clearly in Fig.\ \ref{chi-T-eps-8}(b), which plots the residual
$\chi^{-1}(T)-f(T)$ after subtraction of the fitted leading term. The residual
is seen to be well-described by a power law over about four decades of
temperature $10^{-6}<T<10^{-2}$. The fitted exponent is different from the
value $x_2=\half$ that would correspond to the dashed line shown in
Fig.\ \ref{chi-T-eps-8}(b).

Similar results hold
for $\epsilon=0.7$, fitted over $10^{-5}<T<10^{-1}$ by $x_2=0.671$
[Fig.\ \ref{chi-T-eps-7}(b)];
for $\epsilon=0.6$, fitted over $10^{-5}<T<0.5$ by $x_2=0.665$
[Fig.\ \ref{chi-T-eps-6}(b)];
and for $\epsilon=0.2$, fitted over $5 \times 10^{-5}<T<0.5$ by $x_2=1.25$
[Fig.\ \ref{chi-T-eps-2}(b)].
For each of the three cases in the range $\half<\epsilon<1$ where the
validity of the quantum-to-classical mapping is at issue, we find no
evidence for the $T^{1/2}$ contribution to $\chi^{-1}(T)$ that is
assumed in Ref.\ \onlinecite{Vojta.10}.

\subsection{Frequency dependence of the critical local susceptibility}
\label{chiw_qcp}

\begin{figure}
\centering
\includegraphics[width=0.75\linewidth]{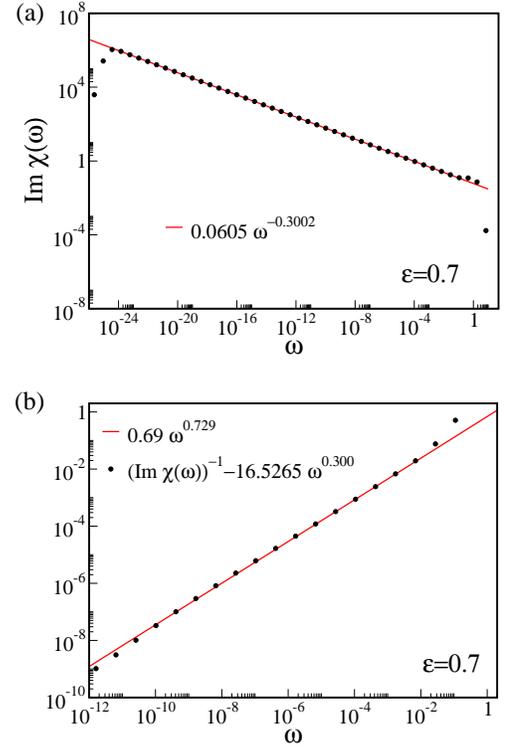}
\caption{(Color online) Frequency dependence of the local dynamical
susceptibility at the QCP ($T=0$, $g=g_c$). (a) $\Im \chi(\omega)$ for
$\epsilon=0.7$, $\Gamma=1$, and $\Lambda=4$.
The solid line represents a power-law fit to Eq.\ \eqref{chi-loc-im-cl-dyn},
which yields the exponent $y=0.3002$.
(b) Residual $[\Im\chi(\omega)]^{-1} - b\omega^y$ after subtraction of the
leading term inferred from (a). The fitted solid line gives a subleading
exponent $y_2=0.729$.}
\label{chi-ome-eps-7}
\end{figure}

\begin{figure}
\centering
\includegraphics[width=0.75\linewidth]{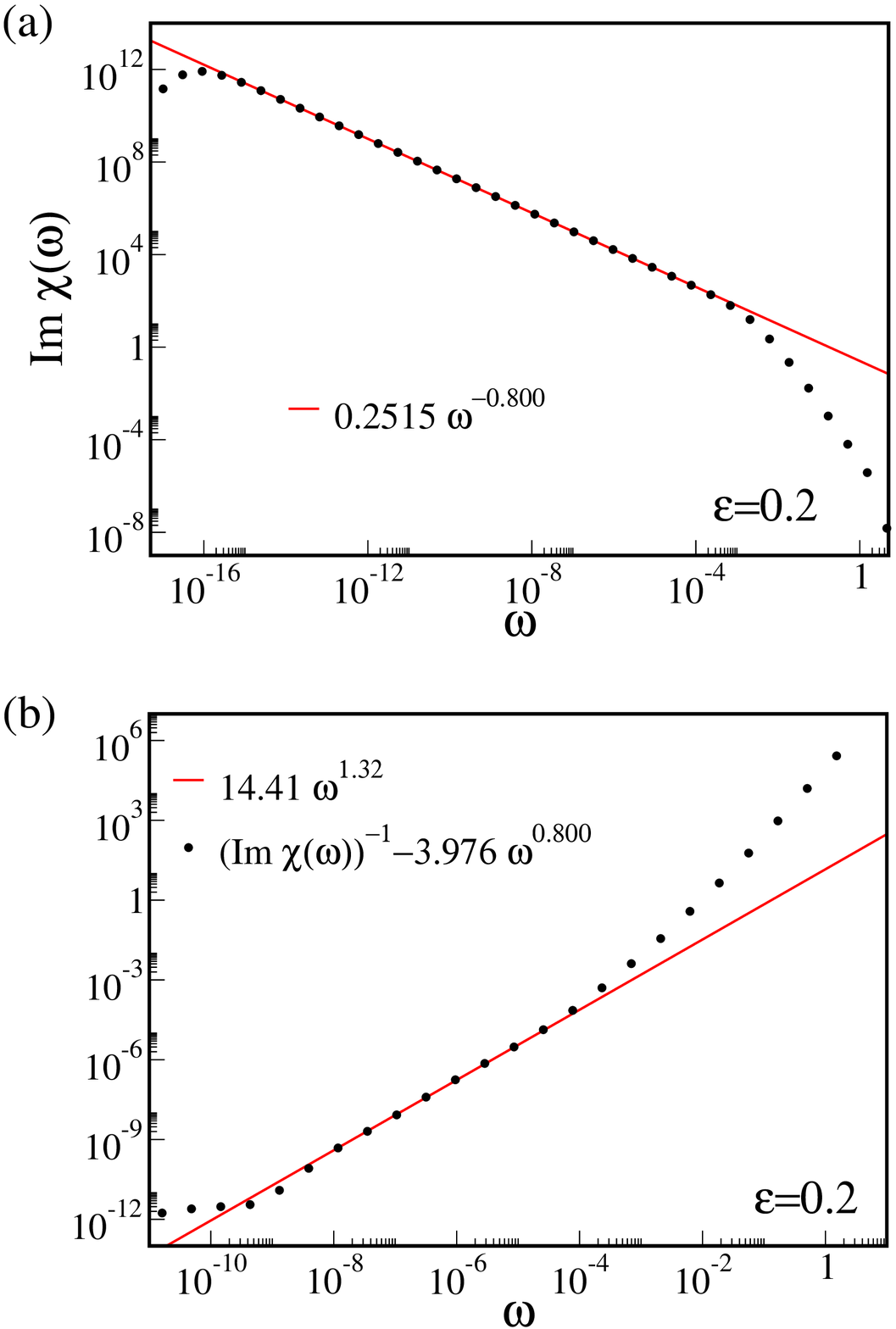}
\caption{(Color online) Like Fig. \ref{chi-ome-eps-7}, but for
$\epsilon=0.2$, $\Gamma=0.01$, and $\Lambda=3$.}
\label{chi-ome-eps-2}
\end{figure}

Next, we investigate the frequency dependence of the zero-temperature critical
local dynamical susceptibility $\chi(\omega)$. Since the NRG delivers
$\chi''(\omega)=\Im\chi(\omega)$, this quantity is the focus of our analysis.
If necessary, $\chi'(\omega)=\Re\chi(\omega)$ can be obtained via a Hilbert
transform of $\chi''(\omega)$, but this procedure introduces some error due
to the lower accuracy of the NRG-calculated $\chi''(\omega)$ at high
frequencies.

Figure \ref{chi-ome-eps-7}(a) plots $\chi''(\omega)$ for $\epsilon=0.7$.
The low-frequency behavior is singular:
\begin{equation}
\chi''(\omega) = \frac{\sgn\,\omega}{b|\omega|^y} \, .
\label{chi-loc-im-cl-dyn}
\end{equation}
The exponent is, to an accuracy considerably better than $1\%$,
$y=1-\epsilon=0.3$. This implies that the full retarded susceptibility 
has the form specified by Eq.\ \eqref{chi-loc-qu-dyn}.
The equality of $y$ and the temperature exponent $x$ is consistent with an
$\omega/T$ scaling form for the leading scaling term in the inverse local
susceptibility.

We will assume that $\chi(\omega)$ has a subleading term with exponent $y_2$
defined through
\begin{equation}
\frac{1}{\chi (\omega) }= B (-i\omega)^y + B_2 (-i\omega)^{y_2} .
\label{chi-ome-fit-leading-and-subleading}
\end{equation}
Correspondingly, for $\omega>0$,
\begin{equation}
\frac{1}{\chi''(\omega)} \simeq b \omega^{y} + b_2 \omega^{y_2} ,
\label{chi-ome-im-fit-leading-and-subleading}
\end{equation}
where $b = B/\sin (\pi y/2)$,
\begin{equation}
b_2 = B_2 \left[ \frac{\cos\left( \pi (y + y_2)/2 \right)}{\sin (\pi y /2)} +
\frac{\sin (\pi y_2 /2) }{\sin (\pi y /2)^2} \right] ,
\end{equation}
and the approximation in Eq.\ \eqref{chi-ome-im-fit-leading-and-subleading}
involves ignoring higher-order terms $\sim \omega ^{2 y_2-y}$.
Fitting to Eq.\ \eqref{chi-ome-im-fit-leading-and-subleading} [Fig.\
\ref{chi-ome-eps-7}(b)] yields an exponent $y_2=0.73$ that agrees
with the subleading temperature exponent $x_2=0.67$ [Fig.\
\ref{chi-T-eps-7}(d)] to within about 9\%. Given that extracting $y_2$
from $1/\chi''(\omega)$ is less accurate than determining the exponent from
$1/\chi(\omega)$ [whose determination requires a calculation of $\chi'(\omega)$,
however], we interpret our results as being consistent with $y_2=x_2$.
This suggests that the subleading term of the inverse 
local spin susceptibility also obeys $\omega/T$ scaling.

\begin{figure}
\centering
\includegraphics[width=0.75\linewidth]{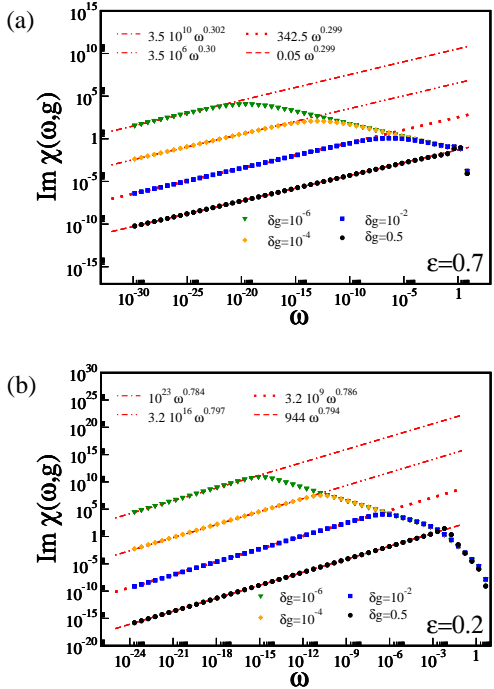}
\caption{(Color online) Imaginary part $\Im\chi(\omega)$ of the
zero-temperature local dynamical susceptibility at four different bosonic
couplings $g < g_c$ or, equivalently, $\delta g \equiv (g_c - g)/g_c > 0$,
for (a) $\epsilon=0.7$, $\Gamma=1$, and $\Lambda=4$;
and (b) $\epsilon=0.2$, $\Gamma=0.01$, and $\Lambda=3$.}
\label{chi-ome-deloc-eps-7-2}
\end{figure}

The same procedure can also be applied to the case $\epsilon=0.2$, as shown
in Figs.~\ref{chi-ome-eps-2}(a) and \ref{chi-ome-eps-2}(b). The extracted
$y$ is again to a good accuracy equal to $1-\epsilon=0.8$, the same as the
leading temperature exponent $x$, so the leading term of the inverse local
susceptibility is consistent with $\omega/T$ scaling. The fitted value of the
subleading frequency exponent is $y_2=1.32$, which agrees with the subleading
temperature exponent $x_2$ to within 6\%. It is worth noting that for this
value of $\epsilon$, which falls in the range $0<\epsilon<\half$, the level of
agreement between $x_2$ and $y_2$ is similar in percentage terms to that for
$\epsilon=0.7$ in the range $\half<\epsilon<1$.

\begin{figure}
\centering
\includegraphics[width=0.8\linewidth]{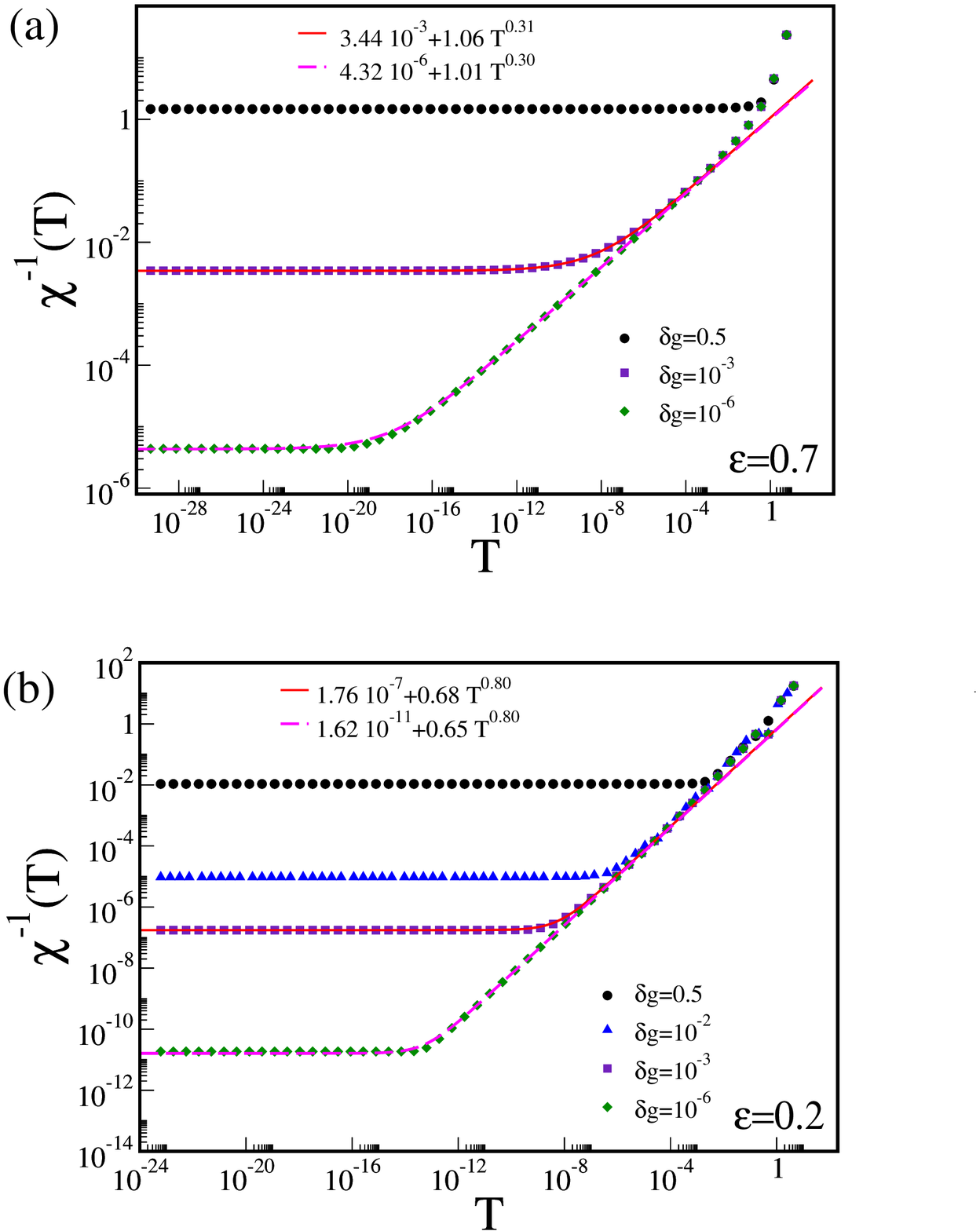}
\caption{(Color online) (a) Inverse $\chi^{-1}(T)$ of the local static
susceptibility at different bosonic couplings $g < g_c$ or, equivalently,
$\delta g \equiv (g_c - g)/g_c > 0$,
for (a) $\epsilon=0.7$, $\Gamma=1$, and $\Lambda=4$;
and (b) $\epsilon=0.2$, $\Gamma=0.01$, and $\Lambda=3$.}
\label{chi-T-deloc-eps-7-2}
\end{figure}

\section{Local correlation functions in the delocalized phase}
\label{chi_deloc}

To further analyze the NRG results, we turn to the local susceptibility in
the delocalized phase, $g<g_c$, which has not received much attention in
previous studies. Figure \ref{chi-ome-deloc-eps-7-2}(a) shows
the $\omega$ dependence of $\chi''(\omega)$ at zero temperature for
$\epsilon=0.7$ and several values of $\delta g \equiv (g_c-g)/g_c$. The
leading frequency dependence is seen to be \
$\chi ''(\omega) \sim |\omega|^{1-\epsilon}$. The results are
well described by
\begin{equation}
\chi(T=0, \omega, g < g_c) = \frac{1}{A + B (-i\omega)^{1-\epsilon}}.
\label{chi-omega-g-lt-gc}
\end{equation}
The decrease of $\chi(\omega=0)$ with increasing $\delta g$ ({\it i.e.},
decreasing $g$) in Fig.~\ref{chi-ome-deloc-eps-7-2}(a) reflects the
increase of $A$ as $g$ moves away from $g_c$.
These conclusions are also valid for $\epsilon=0.2$, as seen in Fig.\
\ref{chi-ome-deloc-eps-7-2}(b).

Figure \ref{chi-T-deloc-eps-7-2}(a) shows that for $\epsilon=0.7$, the
local static susceptibility in the delocalized phase is well described by
\begin{equation}
\chi (T, \omega =0, g < g_c) = \frac{1}{A + a T^{1-\epsilon}} .
\label{chi-T-g-lt-gc}
\end{equation}
The same conclusions, once again, apply to $\epsilon=0.2$, as seen in 
Fig.\ \ref{chi-T-deloc-eps-7-2}(b).

We close this section with two general remarks. First, the exponent
$1-\epsilon$ describing the frequency dependence (and, by extension, the
temperature dependence), takes the largest possible value that satisfies 
the Griffiths inequality,\cite{Griffiths} which states that correlation
functions decay in time no faster than does the interaction [as specified in
this case by Eqs.\ \eqref{Seff} and \eqref{chiInv} below].
Second, this frequency and temperature dependence only exists in the
delocalized phase for $g \ne 0$, implying that turning on the coupling $g$
to the bosonic bath within the delocalized phase is a singular perturbation.

\section{Discussion}
\label{discussion}

\subsection{Implications of our results}

We have considered in detail the critical behavior of the local spin
susceptibility of the sub-ohmic spin-boson model calculated using the NRG.
For four different bath exponents spanning the range $0.2\le\epsilon\le 0.8$,
we have confirmed that the leading term in the inverse static local spin
susceptibility has the $T^{1-\epsilon}$ temperature dependence described by
Eq.\ \eqref{chi-loc-qu} and demonstrated that the subleading term varies as
$T^{x_2}$ with $x_2>\half$. 

This analysis allows us to assess the alternative interpretations (as outlined
in the Introduction) of the Monte-Carlo results for the critical behavior of
the local susceptibility in a long-ranged classical Ising chain. In the regime
$\half<\epsilon < 1$, the chain yields a temperature exponent of $\half$. This
originates from a dangerously irrelevant variable, which also leads to a
violation of $\omega/T$ scaling. In particular, the $T^{\half}$ dependence of
of the inverse local susceptibility is not accompanied by a
$(-i\omega)^{\half}$ dependence on frequency.
These results differ from those of NRG calculations for the quantum model.

The NRG results presented in Sec.\ \ref{chi_qcp} are consistent with the
interpretation advanced in Ref.~\onlinecite{Kirchner.09} in the context
of the Bose-Fermi Kondo model, namely the difference of the Monte-Carlo
results from the NRG results and reflects a violation of the
quantum-to-classical mapping. The latter is the result of a Berry-phase
term, which we discuss in the next subsection. 

Our results are inconsistent with the interpretation put forward in
Refs.\ \onlinecite{Winter.09}, \onlinecite{Vojta.09} and \onlinecite{Vojta.10},
which advocates the validity of quantum-to-classical mapping. In this
picture, the intrinsic temperature dependence of the inverse critical
local susceptibility of the NRG calculation should be $T^{\half}$, which 
is masked by an artificial leading $T^{1-\epsilon}$ term. We have shown
here that there is no subleading $T^{\half}$ term in the NRG results.

\begin{figure}
\centering
\includegraphics[width=0.7\linewidth]{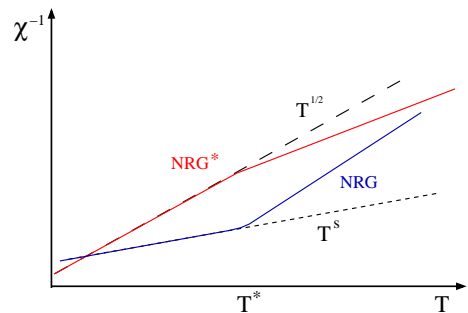}
\caption{Schematic log-log plot of the inverse static local
susceptibility $\chi^{-1}$ vs temperature as obtained using
the NRG (\textit{cf.\ } Figs.~\ref{chi-T-eps-8}--\ref{chi-T-eps-2}
of the present paper) and NRG$^\ast$ [\textit{cf.\ } Figs.\ 7(b)
and 8(b) of Ref.\ \cite{Vojta.10}] procedures. The curvature of
the NRG$^{\ast}$ line suggests  that a $T^{1/2}$ dependence at
$T < T^{\ast}$ crosses over at $T>T^{\ast}$ to a $T^{z}$
dependence with a $z$ \textit{smaller} than $\half$; this in turn
suggests that the asymptotic low-temperature behavior has a
temperature exponent less than $\half$.}
\label{fig:nrg-nrg-star}
\end{figure}

Reference \onlinecite{Vojta.10} suggested the possibility of removing 
the leading $T^{1-\epsilon}$ term using an ``NRG$^{\ast}$'' procedure
involving NRG calculations for the spin-boson model supplemented by
an \textit{ad hoc} Hamiltonian term containing a coefficient that was
adjusted until the inverse local susceptibility at the QCP had a
temperature exponent close to \half.
Figure \ref{fig:nrg-nrg-star} provides a schematic comparison between the
temperature dependences of the critical static local susceptibilities
produced by the NRG (our data from Sec.\ \ref{chiT_qcp} above) and by the
NRG$^{\ast}$ procedure (Figs.\ 7 and 8 of Ref.\ ~\onlinecite{Vojta.10}).
If the leading term in $\chi^{-1}(T)$ corresponded to $x=\half$, then
corrections arising from any subleading term with an exponent $z>\half$
would induce a crossover in $\chi^{-1}(T)$ around some $T^\ast$ to a
higher-temperature behavior with a \textit{greater} slope. Instead, the
$\epsilon=0.6$ and $0.7$ NRG$^{\ast}$ results of Ref.~\onlinecite{Vojta.10}
appear to show a $T^z$ dependence with $z<\half$ for $T>T^\ast$.
One should expect, based on criticality, any such $T^z$ term to persist
to sufficiently low temperatures that it, not the $T^{\half}$ term,
dominates the asymptotic behavior.

The $T^{x_2}$ subleading term in the temperature dependence of the critical
inverse static susceptibility is accompanied by a subleading term in the
frequency dependence of the inverse dynamical susceptibility of the form
$(-i\omega)^{y_2}$. Our results are compatible with $y_2=x_2$ and, in turn,
with $\omega/T$ scaling for the subleading term (as well as the leading term,
where $y=x$). If one interprets both the $T^x$ and $T^{x_2}$ terms in the
temperature dependence as artifacts of NRG, then any intrinsic term must vary
with a power of $T$ even greater than $x_2$; since $x_2>\half$, such behavior
would be inconsistent with the validity of the quantum-to-classical mapping.
Instead, the result reinforces the conclusion that the fixed point is
interacting and features $\omega/T$ scaling.

We now briefly discuss other critical exponents describing the
variation of the local magnetization
$M=\langle S^z\rangle$, the
corresponding susceptibility $\chi=\left.\partial M/
\partial h\right|_{h=0}$, and the (imaginary-time) correlation
length $\xi_{\tau}$ in the vicinity of the QCP:
\begin{align}
M(T=0,h=0,g > g_c) &\propto (g-g_c)^{\beta}, \\ 
M(T=0,h,g=g_c) &\propto |h|^{1/\delta}, \\
\chi(T=0,g<g_c) &\propto (g_c-g)^{-\gamma}, \\
\xi_{\tau} &\propto |g - g_c|^{-\nu} .
\end{align}
An important conclusion from the NRG treatment of the
spin-boson model\cite{Vojta.05} (as well as that of the
Bose-Fermi Kondo model\cite{Glossop.07}) is that,
together with $x$ and $y$, these exponents satisfy hyperscaling
relations \cite{Ingersent.02} expected to hold at an interacting QCP.
Ref.~\onlinecite{Vojta.10} argued that the NRG results for $\delta$ and
$\beta$ are invalid for a reason that is entirely separate from
mass-flow error, namely the NRG truncation of the bosonic Hilbert space.
(A variational procedure has recently been proposed\cite{Hou.10,Guo.11}
to circumvent the effects of bosonic truncation.)
However, it appears unnatural that two independent sources of errors 
should conspire to yield results that are consistent with hyperscaling.

\subsection{Path integral and Berry phase}

Any proper path integral requires taking the continuum limit in time.
The quantum critical properties of the Bose-Fermi Kondo model or the
spin-boson model can be described within a proper path integral
representation. It was shown in Ref.\ \onlinecite{Kirchner.08e} that the
hyperscaling observed in the large-$N$ limit of the
$\mathrm{SU}(N)\times\mathrm{SU}(M)$ Bose-Fermi Kondo model even for
$\half<\epsilon<1$ can be understood as a consequence of a topological
Berry-phase term that characterizes the path integral over $\mathrm{SU}(N)$.
Thus, the breakdown of the quantum-to-classical mapping is not merely an
artifact of the saddle point approximation, and it survives the inclusion
of $1/N$ corrections.\cite{Kirchner.08e} The path integral over the group
$\mathrm{SU}(2)$ also involves a Berry-phase term that spoils the
reinterpretation of the resulting complex action in terms of a classical
action. In other words, the quantum-to-classical mapping cannot be
applied for $N=2$ and the quantum critical properties have to be obtained
directly from the quantum model. A proper path integral on $\mathrm{SU}(2)$
is given by the functional integral on the Bloch sphere $S^2$, the coset
space of $\mathrm{SU}(2)$. Every path on $S^2$ is characterized by a
geometric phase independent of the spin Hamiltonian. It therefore is
natural to expect similar spin path integral representations for the
$\mathrm{SU}(2)$ Bose-Fermi Kondo model and for the spin-boson or easy-axis
Bose-Fermi Kondo model.

It was observed in Ref.\ \onlinecite{Kirchner.10} that the construction
of a Feynman path integral for the spin-boson model through a
time-slicing procedure encounters difficulties in taking the continuum limit
when using the orthonormal eigenfunction basis of $S^z$.
The standard procedure for circumventing this issue in the case of pure
spin Hamiltonians is to rewrite the short-time propagator as a transfer
matrix in spin space.\cite{Suzuki.76} A generalization of this method
to the spin-boson model rewrites the matrix element of the infinitesimal
(imaginary) time evolution operator as

\begin{equation}
\langle\sigma_i \phi_i|e^{-H\tau_0}|\sigma_{i+1}\phi_{i+1}\rangle =
 e^{\phi_i^* \phi_{i+1}^{}} \bm{P}^{i,i+1},
\end{equation}
where $\bm{P}^{i,i+1}$ is a matrix in the spin subspace, $\phi$ is a
c-number that labels the bosonic coherent states,
$e^{\phi_i^* \phi_{i+1}^{}}$ is related to the bosonic Berry phase,
and $i$ indexes the time slices.
The matrix elements of ${\bm P}^{i,i+1}$ are then expressed as the exponential of
some function $F(S_i,S_{i+1})$ with $S_i=\pm 1$:\cite{Kirchner.10}

\begin{equation}
\label{versuch2}
\bm{P}^{i,i+1}\Big|_{S_i,S_{i+1}}=e^{a+b(S_i+S_{i+1})+c S_i S_{i+1}},
\end{equation}
with
\begin{align*}
a&=\half\bigl[-\omega\tau_0\,\phi^*_{i+1}\phi^{}_i + \ln(\Gamma\tau_0)\bigr],\\
b&=-\half g\tau_0 \bigl(\phi^*_{i+1}+\phi^{}_i \bigr),\\
c&=\half\bigl[-\omega\tau_0\,\phi^*_{i+1}\phi^{}_i - \ln(\Gamma\tau_0)\bigr].
\end{align*}
Inserting Eq.\ \eqref{versuch2} into the expression for the partition function

\begin{align}
\label{partition1}
\mathcal{Z}
&= \mbox{Tr} \, e^{-\beta H} \\
&= A\prod_{i=0}^M \sum_{\sigma_i=\uparrow,\downarrow}
 \int d\phi_i^* d\phi_i^{} \, e^{-\phi_i^* \phi_i^{}}
 \langle\sigma_M \phi_M |e^{-H\tau_0}| \sigma_0 \phi_0\rangle \notag\\
&\qquad\times \prod_{k=0}^{M-1}
 \langle\sigma_k \phi_k |e^{-H\tau_0}|\sigma_{k+1} \phi_{k+1} \rangle \notag
\end{align}
(where $A$ is a constant and $\tau_0=\beta/M$), results in an effective 
action having the form of the action for a classical spin chain coupled
to the $\phi$-fields with nearest-neighbor interaction 
$\half\bigl[-\omega\tau_0\,\phi^{*}_{i+1}\phi^{}_i-\ln(\Gamma\tau_0)\bigr]$,
which is singular as $\tau_0 \rightarrow 0$.
However, the terms coupling the bosons are not singular for
$\tau_0\rightarrow 0$, making it difficult to regularize this limit.
Nonetheless, keeping track of both the transverse magnetic field $\Gamma$ and
the coupling $g$ to the bosonic bath is essential given that the competition
between these two yields the QCP, and also that turning on $g$ is a singular
perturbation (as discussed at the end of Sec.\ \ref{chi_deloc}).
A related way to see this difficulty is that, for $\tau_0 \rightarrow 0$,
the Kondo-like energy scale goes to zero and the topological effect encoded
in the spin flips is suppressed.\cite{Kirchner.09}

On the other hand, in a spin coherent-state representation of the
spin-boson model, the continuum limit following a time-slicing procedure
poses no difficulty. It leads to a well-defined path integral representation
for the sub-ohmic spin-boson model
[Eq.\ \eqref{EQ:sub-Ohmic}]:\cite{Kirchner.10}
\begin{equation}
\mathcal{Z}=\int\mathcal{D}[\vec{n}]\, \exp[-S_{\mathrm{eff}}/2],
\label{eq:SBpartition}
\end{equation}
where the effective action for the spin degrees of freedom is
\begin{align}
\label{Seff}
S_{\mathrm{eff}}
&= -i\mathcal{A}[\vec{n}] -\!\,\! \int_0^\beta d\tau \Gamma  n_x(\tau)\\
&+\frac{g^2}{2} \int_0^{\beta} \! d\tau \int_0^{\beta} \! d\tau' \,
 n_z(\tau) \, \chi^{-1}_0(\tau-\tau') \, n_z(\tau') , \notag
\end{align}
where\cite{Kirchner.09}
\begin{equation}
\label{chiInv}
\chi^{-1}_0(\tau-\tau^{'})=\int_0^{\infty}\!d\omega \, \omega^{1-\epsilon}\,
\frac{\cosh[\omega(\beta/2-|\tau-\tau^{'}|)]}{\sinh(\omega\beta/2)},
\end{equation}
and $\vec{n}$ defines a point on $S^2$ ($|\vec{n}(\tau)|^2=1$).
The effective action of the sub-ohmic spin-boson problem in the continuum
limit contains the Berry-phase term $-i\mathcal{A}[\vec{n}]$, where
$\mathcal{A}[\vec{n}]$ is equivalent to the area on the sphere $S^2$ traced out
by $\vec{n}(\tau)$ for $0\leq \tau\leq \beta$ with $\vec{n}(0)=\vec{n}(\beta)$.
As discussed previously for models with continuous spin
symmetry\cite{Kirchner.08e}, the Berry-phase term in the action,
being imaginary, can invalidate the mapping of the quantum action to a
classical one.

\section{Summary}
\label{summary}

We have revisited the critical behavior of the sub-ohmic spin-boson problem.
An analysis of the subleading term in the temperature dependence of the inverse
local spin susceptibility at the quantum critical point as calculated using
the numerical renormalization-group (NRG) method has provided evidence
that the instrinsic behavior of the critical static spin susceptibility is
given by Eq.\ \ref{chi-loc-qu} with a non-mean-field exponent. More
specifically, we have examined the implications of our data for the two
interpretations of the difference (for $\half<\epsilon<1$) between Monte-Carlo
calculations for a classical Ising spin-chain and NRG calculations for the
quantum spin-boson model. The leading and subleading temperature dependences
of the critical local susceptibility are consistent only with the
interpretation that the classical spin-chain model fails to capture the
critical behavior of the quantum model and that the criticality is described
by an interacting fixed point characterized by $\omega/T$ scaling.
This conclusion is further supported by the fact that the subleading frequency
dependence of the inverse critical dynamical spin susceptibility has (within
numerical uncertainty) the same exponent as that for the subleading temperature
dependence of the inverse critical static spin susceptibility.

These results provide evidence for the violation of the quantum-to-classical
mapping in the sub-ohmic spin-boson model.
We have discussed the effect of a Berry-phase term in a continuum path-integral
representation for this model. The importance of the Berry-phase term connects 
the violation of the quantum-to-classical mapping observed here with that seen
in models with continuous spin symmetry.

We thank R.\ Bulla, J.\ von Delft, M.\ Glossop, H.\ Rieger, M.\ Troyer, M.\ Vojta, and T.\ Vojta 
for useful discussions. This work has been supported by
NSF Grants No.\ DMR-0710540, DMR-1006985, and DMR-1107814,
and by Robert A.\ Welch Foundation Grant No.\ C-1411.


\begin{thebibliography}{16}
\expandafter\ifx\csname natexlab\endcsname\relax\def\natexlab#1{#1}\fi
\expandafter\ifx\csname bibnamefont\endcsname\relax
  \def\bibnamefont#1{#1}\fi
\expandafter\ifx\csname bibfnamefont\endcsname\relax
  \def\bibfnamefont#1{#1}\fi
\expandafter\ifx\csname citenamefont\endcsname\relax
  \def\citenamefont#1{#1}\fi
\expandafter\ifx\csname url\endcsname\relax
  \def\url#1{\texttt{#1}}\fi
\expandafter\ifx\csname urlprefix\endcsname\relax\def\urlprefix{URL }\fi
\providecommand{\bibinfo}[2]{#2}
\providecommand{\eprint}[2][]{\url{#2}}

\bibitem{Hewson}
A.~C.~Hewson,
\textit{The {K}ondo {P}roblem to {H}eavy {F}ermions}
(Cambridge University Press, 1993).

\bibitem[{\citenamefont{Zhu and Si}(2002)}]{Zhu.02}
\bibinfo{author}{\bibfnamefont{L.}~\bibnamefont{Zhu}} \bibnamefont{and}
  \bibinfo{author}{\bibfnamefont{Q.}~\bibnamefont{Si}},
  \bibinfo{journal}{Phys.~Rev.~B} \textbf{\bibinfo{volume}{66}},
  \bibinfo{pages}{024426} (\bibinfo{year}{2002}).

\bibitem{Zarand.02}
  G.\ Zar\'{a}nd and E.\ Demler, Phys.\ Rev.\ B \textbf{66}, 024427 (2002).

\bibitem[{\citenamefont{Glossop and Ingersent}(2007)}]{Glossop.07}
\bibinfo{author}{\bibfnamefont{M.}~\bibnamefont{Glossop}} \bibnamefont{and}
  \bibinfo{author}{\bibfnamefont{K.}~\bibnamefont{Ingersent}},
  \bibinfo{journal}{Phys.~Rev.~B} \textbf{\bibinfo{volume}{75}},
  \bibinfo{pages}{104410} (\bibinfo{year}{2007}).

\bibitem[{\citenamefont{Kirchner et~al.}(2009)\citenamefont{Kirchner, Si, and
  Ingersent}}]{Kirchner.09}
\bibinfo{author}{\bibfnamefont{S.}~\bibnamefont{Kirchner}},
  \bibinfo{author}{\bibfnamefont{Q.}~\bibnamefont{Si}}, \bibnamefont{and}
  \bibinfo{author}{\bibfnamefont{K.}~\bibnamefont{Ingersent}},
  \bibinfo{journal}{Phys.~Rev.~Lett.} \textbf{\bibinfo{volume}{102}},
  \bibinfo{pages}{166405} (\bibinfo{year}{2009}).

\bibitem[{\citenamefont{Zhu et~al.}(2004)\citenamefont{Zhu, Kirchner, Si, and
  Georges}}]{Zhu.04}
\bibinfo{author}{\bibfnamefont{L.}~\bibnamefont{Zhu}},
  \bibinfo{author}{\bibfnamefont{S.}~\bibnamefont{Kirchner}},
  \bibinfo{author}{\bibfnamefont{Q.}~\bibnamefont{Si}}, \bibnamefont{and}
  \bibinfo{author}{\bibfnamefont{A.}~\bibnamefont{Georges}},
  \bibinfo{journal}{Phys.~Rev.~Lett.} \textbf{\bibinfo{volume}{93}},
  \bibinfo{pages}{267201} (\bibinfo{year}{2004}).

\bibitem[{\citenamefont{Fisher et~al.}(1972)\citenamefont{Fisher, Ma, and
  Nickel}}]{Fisher.72}
\bibinfo{author}{\bibfnamefont{M.~E.} \bibnamefont{Fisher}},
  \bibinfo{author}{\bibfnamefont{S.}~\bibnamefont{Ma}}, \bibnamefont{and}
  \bibinfo{author}{\bibfnamefont{B.~G.} \bibnamefont{Nickel}},
  \bibinfo{journal}{Phys.~Rev.~Lett.} \textbf{\bibinfo{volume}{29}},
  \bibinfo{pages}{917} (\bibinfo{year}{1972}).

\bibitem[{\citenamefont{Kirchner and Si}()}]{Kirchner.08e}
\bibinfo{author}{\bibfnamefont{S.}~\bibnamefont{Kirchner}} \bibnamefont{and}
  \bibinfo{author}{\bibfnamefont{Q.}~\bibnamefont{Si}},
  \bibinfo{note}{arXiv:0808.2647 (2008)}.

\bibitem{Vojta.05}
  M.\ Vojta, N.\ H.\ Tong, and R.\ Bulla,
  Phys.\ Rev.\ Lett.\ \textbf{94}, 070604 (2005).

\bibitem[{\citenamefont{Guinea et~al.}(1985)\citenamefont{Guinea, Hakim, and
  Muramatsu}}]{Guinea.85}
\bibinfo{author}{\bibfnamefont{F.}~\bibnamefont{Guinea}},
  \bibinfo{author}{\bibfnamefont{V.}~\bibnamefont{Hakim}}, \bibnamefont{and}
  \bibinfo{author}{\bibfnamefont{A.}~\bibnamefont{Muramatsu}},
  \bibinfo{journal}{Phys.~Rev.~B} \textbf{\bibinfo{volume}{32}},
  \bibinfo{pages}{4410} (\bibinfo{year}{1985}).

\bibitem{Bulla.03}
  R.\ Bulla, N.\ H.\ Tong, and M.\ Vojta,
  Phys.\ Rev.\ Lett.\ \textbf{91}, 170601 (2003).

\bibitem[{\citenamefont{Bulla et~al.}(2005)\citenamefont{Bulla, Lee, Tong, and
  Vojta}}]{Bulla.05}
\bibinfo{author}{\bibfnamefont{R.}~\bibnamefont{Bulla}},
  \bibinfo{author}{\bibfnamefont{H.}~\bibnamefont{Lee}},
  \bibinfo{author}{\bibfnamefont{N.}~\bibnamefont{Tong}}, \bibnamefont{and}
  \bibinfo{author}{\bibfnamefont{M.}~\bibnamefont{Vojta}},
  \bibinfo{journal}{Phys.~Rev.~B} \textbf{\bibinfo{volume}{71}},
  \bibinfo{pages}{045122} (\bibinfo{year}{2005}).

\bibitem[{\citenamefont{Winter et~al.}(2009)\citenamefont{Winter, Rieger,
  Vojta, and Bulla}}]{Winter.09}
\bibinfo{author}{\bibfnamefont{A.}~\bibnamefont{Winter}},
  \bibinfo{author}{\bibfnamefont{H.}~\bibnamefont{Rieger}},
  \bibinfo{author}{\bibfnamefont{M.}~\bibnamefont{Vojta}}, \bibnamefont{and}
  \bibinfo{author}{\bibfnamefont{R.}~\bibnamefont{Bulla}},
  \bibinfo{journal}{Phys.~Rev.~Lett.} \textbf{\bibinfo{volume}{102}},
  \bibinfo{pages}{030601} (\bibinfo{year}{2009}).

\bibitem[{\citenamefont{Luijten and Bl\"{o}te}(1997)}]{Luijten.97}
\bibinfo{author}{\bibfnamefont{E.}~\bibnamefont{Luijten}} \bibnamefont{and}
  \bibinfo{author}{\bibfnamefont{H.~W.~J.} \bibnamefont{Bl\"{o}te}},
  \bibinfo{journal}{Phys.~Rev.~B.} \textbf{\bibinfo{volume}{56}},
  \bibinfo{pages}{8945} (\bibinfo{year}{1997}).

\bibitem[{\citenamefont{Br{\'e}zin}(1982)}]{Brezin.82}
\bibinfo{author}{\bibfnamefont{E.}~\bibnamefont{Br{\'e}zin}},
  \bibinfo{journal}{J.~Physique} \textbf{\bibinfo{volume}{43}},
  \bibinfo{pages}{15} (\bibinfo{year}{1982}).

\bibitem[{\citenamefont{Kirchner}(2010)}]{Kirchner.10}
\bibinfo{author}{\bibfnamefont{S.}~\bibnamefont{Kirchner}},
  \bibinfo{journal}{J.~Low Temp.~Phys.} \textbf{\bibinfo{volume}{161}},
  \bibinfo{pages}{282} (\bibinfo{year}{2010}).

\bibitem{Vojta.09}
  M.\ Vojta, N.\ H.\ Tong, and R.\ Bulla,
  Phys.\ Rev.\ Lett.\ \textbf{102}, 249904(E) (2009).
  
\bibitem[{\citenamefont{Vojta et~al.}(2010)\citenamefont{Vojta, Bulla,
  G\"uttge, and Anders}}]{Vojta.10}
\bibinfo{author}{\bibfnamefont{M.}~\bibnamefont{Vojta}},
  \bibinfo{author}{\bibfnamefont{R.}~\bibnamefont{Bulla}},
  \bibinfo{author}{\bibfnamefont{F.}~\bibnamefont{G\"uttge}}, \bibnamefont{and}
  \bibinfo{author}{\bibfnamefont{F.}~\bibnamefont{Anders}},
  \bibinfo{journal}{Phys.~Rev.~B} \textbf{\bibinfo{volume}{81}},
  \bibinfo{pages}{075122} (\bibinfo{year}{2010}).
  
\bibitem{note_mass_winding}
The mass term proportional to $T^{1-\epsilon}$ introduced in 
Ref.\ \onlinecite{Kirchner.09} for the classical Ising chain of length
$\propto 1/T$ is associated with a change of the interaction potential.
This change corresponds to a truncation of ``winding" terms of
the long-ranged interaction around the periodic chain.
  
\bibitem{Griffiths}
R.\ Griffiths, J.\ Math.\ Phys.\ \textbf{8}, 478 (1970).

\bibitem{Ingersent.02}
K.\ Ingersent and Q.\ Si,
Phys.\ Rev.\ Lett.\ \textbf{89}, 076403 (2002).

\bibitem[{\citenamefont{Hou and Tong}(2010)}]{Hou.10}
\bibinfo{author}{\bibfnamefont{Y.-H.} \bibnamefont{Hou}} \bibnamefont{and}
  \bibinfo{author}{\bibfnamefont{N.-H.} \bibnamefont{Tong}},
  \bibinfo{journal}{Eur.~Phys.~J.~B} \textbf{\bibinfo{volume}{78}},
  \bibinfo{pages}{127} (\bibinfo{year}{2010}).

\bibitem[{\citenamefont{Guo et~al.}(2011)\citenamefont{Guo, Weichselbaum,
  v.~Delft, and Vojta}}]{Guo.11}
\bibinfo{author}{\bibfnamefont{C.}~\bibnamefont{Guo}},
  \bibinfo{author}{\bibfnamefont{A.}~\bibnamefont{Weichselbaum}},
  \bibinfo{author}{\bibfnamefont{J.}~\bibnamefont{von~Delft}}, \bibnamefont{and}
  \bibinfo{author}{\bibfnamefont{M.}~\bibnamefont{Vojta}}
  (\bibinfo{year}{2011}), \bibinfo{note}{arXiv:1110.6314}.

\bibitem[{\citenamefont{Suzuki}(1976)}]{Suzuki.76}
\bibinfo{author}{\bibfnamefont{M.}~\bibnamefont{Suzuki}},
  \bibinfo{journal}{Prog.~Theor.~Phys.} \textbf{\bibinfo{volume}{56}},
  \bibinfo{pages}{1454} (\bibinfo{year}{1976}).
\end{thebibliography}
\end{document}